# Crystal structure of synthetic $Mg_3Cr_2Si_3O_{12}$, the high-pressure Cr end-member of the knorringite-pyrope garnet series

Amélie Juhin [1], Guillaume Morin [1], Erik Elkaim [2], Dan J. Frost [3], Farid Juillot [1], Georges Calas [1]

[1] Institut de Minéralogie et de Physique des Milieux Condensés, UMR CNRS 7590 / Université Paris 6 / Université Paris 7 / IPGP - 140 rue de Lourmel 75015 Paris, France
[2] Synchrotron SOLEIL, L'Orme des Merisiers, Saint-Aubin - BP 48 91192 Gif-sur-Yvette Cedex, France
[3] Bayerisches Geoinstitut, Universität Bayreuth D-95440 Bayreuth, Germany

## ABSTRACT

Knorringite, the Cr-end-member of the pyrope garnet series (Nixon et al. 1968), often occur in high proportions in kimberlite garnets and is thus used for tracing high-pressure deep-earth conditions favorable to the formation of diamonds, in which knorringite-rich garnet can occur as inclusions. However, although the synthesis of knorringite is reported in the literature (Ringwood 1977; Irifune et al. 1982; Taran et al. 2004), the structure of the pure end-member has not been yet determined from experimental data. In this study, the crystal structure of knorringite, $Mg_3Cr_2(SiO_4)_3$, has been refined from high resolution synchrotron X-ray powder diffraction data recorded under ambient conditions on a polycrystalline sample synthesized at 12 GPa in a multi-anvil apparatus. The structure is cubic, space group $Ia\text{-}3d$, $a$ = 11.5935(1), $V$ = 1558.27(4) Å$^3$, $d_{calc}$ = 3.97 g.cm$^{-3}$. The Cr-O distance of 1.957(2) Å is consistent with EXAFS results on the same sample. This short distance indicates a substantial compression of the $CrO_6$ octahedron, compared to ambient pressure $Cr^{3+}$-minerals such as uvarovite (<Cr-O> = 1.99 Å, Andrut and Wildner 2002). Our experimental results thus confirm early empirical predictions based on series of high-pressure Cr-garnet end-members (Fursenko 1981), showing that the values of the Cr-O distance and the Cr-O-Si angle decrease with the augmentation of pressure and with the diminution of the size of the divalent cation.

# INTRODUCTION

Knorringite $Mg_3Cr_2Si_3O_{12}$, the Cr-end-member of the pyrope garnet series, is an important component of garnets in deeper parts of the upper mantle (Irifune et al. 1982). It often occurs in high proportions in kimberlite garnets: thus, the Cr-concentration is used for tracing high-pressure deep-earth conditions favorable to the formation of diamonds, in which knorringite-rich garnet can occur as inclusions. The highest knorringite content known so far for natural garnets inclusions in diamonds from kimberlites is 66.4 mole % (Stachel and Harris 1997). However, the structure of the knorringite end-member has not been still refined. It has been predicted from the structure of natural garnets using empirical laws involving the radii of the cations (Novak and Gibbs. 1971) and by performing least-square refinement of the interatomic distances (Ottonello et al. 1996). Even more recently, first-principles calculations based on Density Functional Theory have been performed (Milman et al. 2001), but the theoretical structure is in disagreement with the predictions. In particular, the calculated Cr-O distance (1.976 Å) is larger than the two predicted ones (1.958 Å and 1.960 Å).

The remaining uncertainty on the structure of knorringite certainly prevents to complete the investigation of the thermodynamical and structural properties associated with the incorporation of chromium in garnets. For example, the Cr-O distance in Cr-containing pyrope was found to be 1.96 Å (Juhin et al, 2008), but the extension of the structural relaxation (identified as a partial or full process) depends on the Cr-O distance taken for the knorringite end-member.

The synthesis of knorringite is reported by several studies (Ringwood 1977; Irifune et al. 1982; Taran et al. 2004). The space group (*Ia-3d*) and the lattice parameter (11.600(1) Å, Ringwood 1977; 11.596(1) Å, Irifune et al. 1982) have been determined from X-ray diffraction data, but the crystal structure has not been still refined. In this study, we report new results on the crystal structure of knorringite, $Mg_3Cr_2Si_3O_{12}$ obtained from high resolution synchrotron X-ray powder diffraction data on a synthetic sample. The sample was synthesized in a multi-anvil apparatus at P = 12 GPa and T = 1500°C. The Rietveld refined distances are compared to those obtained from Extended X-ay Absorption Fine Structure (EXAFS) measurements performed at the Cr K-edge on the same sample. Finally, the crystal structure of $Mg_3Cr_2Si_3O_{12}$ is compared to that of other Cr-garnet end-members, $Ca_3Cr_2Si_3O_{12}$, $Fe_3Cr_2Si_3O_{12}$ and $Mn_3Cr_2Si_3O_{12}$.

# MATERIALS AND EXPERIMENTAL

**High-pressure synthesis**

Starting materials were stoechiometric mixtures of dried MgO, $Cr_2O_3$ and noncrystalline $SiO_2$. The starting mixtures plus 1 μL of water were encapsulated in Re capsules, which were enclosed in the high-pressure cell of an octahedral multianvil apparatus. The two synthesis runs were performed at the Bayerisches Geoinstitut (University of Bayreuth, Germany), using a 1000-ton press and a 14M sample assembly of tungsten carbide anvils. The samples were heated using a $LaCrO_3$ furnace and the temperature was controlled to ± 1°C using a $W_{97}Re_3$ / $W_{75}Re_{25}$ thermocouple. Run conditions, which are reported in Table 1, are close to those of Taran et al. (2004). The experimental calibrated pressures are given ± 0.3 GPa. The samples were quenched by switching off the power to the furnace, with of quench rate > 200 °C.s$^{-1}$. The run products were examined under a binocular microscope. In order to get a rough estimate of their composition, powder x-ray diffraction patterns were recorded using a Panalytical X'Pert Pro MPD diffractometer with CuKα radiation. In both samples referred to as *Kn-1* and *Kn-2*, knorringite was found to be the major phase, associated to a small (similar) amount of eskolaïte (α-$Cr_2O_3$) and to traces of stishovite ($SiO_2$).

**X-ray powder diffraction using synchrotron radiation**

A high-resolution x-ray powder diffraction-pattern of the synthetic *Kn-1* sample was recorded in 3 h in transmission Debye-Scherrer geometry at the CRISTAL undulator beamline at the SOLEIL synchrotron facility, Saclay, France. Approximately 10 mg of sample were mounted in a rotating silica capillary 300 μm in diameter. The pattern was recorded within the 6 – 60 °2θ range with a wavelength of 0.7285 Å. The diffracted beam was collected using a Ge(111) analyser leading to an instrumental peak width lower than 0.02° 2θ which was evaluated by recording a few Bragg peaks of the $LaB_6$ powder diffraction standard.

**EXAFS experiments**

Approximately 10 mg of sample *Kn-2* were finely grinded together with cellulose, in order to make a pellet. Two Cr K-edge X-ray Absorption spectra were collected at room temperature on beamline BM30b (FAME), at the European Synchrotron Radiation Facility (Grenoble, France) operated at 6 GeV. Calibration was made with respect to the first inflection point in a Cr metal foil (5989 eV). The data were recorded using the fluorescence mode with a Si (220) double crystal and a Canberra 30-element Ge detector (Proux et al. 2006), with a spacing of 0.5 eV and of 0.05 Å$^{-1}$ respectively in the XANES and EXAFS regions.

EXAFS data were corrected for self-absorption and extracted using the XAFS program (Winterer 1997). Radial distribution functions around the Cr absorber were obtained by calculating the Fourier transform (FT) of the $k^3\chi(k)$ EXAFS function using a Kaiser–Bessel window within the 2–15 Å$^{-1}$ $k$-range with a Bessel weight of 2.5. Least-squares fitting of the unfiltered $k^3\chi(k)$ functions was performed with the planewave formalism, using a Levenberg–Marquard minimization algorithm. Theoretical phase-shift and amplitude functions employed in this fitting procedure were calculated with the curved-wave formalism using the FEFF 8 code (Ankudinov et al. 1998) and the knorringite predicted structure (Novak and Gibbs 1971). The fit quality was estimated using a reduced $\chi^2$ of the following form:

$$\chi^2_{FT} = \frac{N_{ind}}{n(N_{ind}-p)}\sum_{i=1}^{n}(\|FT\|_{\exp_i} - \|FT\|_{calc_i})^2,$$

where $N_{ind}$ is the number of independent parameters ($N_{ind} = \frac{2\Delta k \Delta R}{\pi}$), $p$ the number of free fit parameters, $n$ the number of data points fitted, and $\|FT\|_{\exp}$ and $\|FT\|_{calc}$ the experimental and theoretical Fourier transform magnitude within the [0.5-6 Å] $R$-range of the $k^3$-weighted EXAFS signal. The number of allowable independent parameters is 45, and our fit included at most 15 variable parameters.

## RESULTS AND DISCUSSION

**Structure refinement**

The structure was Rietveld refined from the experimental synchrotron X-ray powder-diffraction pattern in the range 6-60° $2\theta$, using the program XND (Berar and Baldinozzi 1998). Absorption for Debye Scherrer geometry (Rouse et al. 1970) was negligible for $\mu r = 0.3$ cm$^{-1}$ (assuming that the apparent density of the powder would be half that of the mineral). Scale factors, cell parameters and width and shape parameters of pseudo-Voigt line-profile functions were refined for knorringite $Mg_3Cr_2Si_3O_{12}$, eskolaite α-$Cr_2O_3$ and stishovite $SiO_2$. Atom parameters of these two latter phases were kept fixed to the literature values (Newnham and deHaan 1962; Ross et al. 1990). Indeed, quantitative analysis derived from scale factor values, using the classical method without internal standard (Bish and Post 1989) indicated that eskolaite and stishovite weight fraction in the sample were 10 ± 2 wt% and ~1 wt%, respectively. For knorringite, all atom positions and displacement parameters were refined, together with site occupancies of the cations, using the predicted structure (Novak and Gibbs 1971) as starting model. About 174 Bragg reflections from knorringite were fit in the 6–60° $2\theta$ range and the $R_{wp}$ parameter for the pattern and the $R_{Bragg}$ parameter for knorringite reached final values of 0.125 and 0.037, respectively (Table 2). Final atom positions, displacement parameters, and site occupancies are reported in Table 3.

The Rietveld refined interatomic distances are similar to those determined from predictions based on least-square refinement procedure in a series of natural samples (Novak and Gibbs 1971; Ottonello et al. 1996): in particular, the Rietveld refined Cr-O distance is equal to 1.957(2) Å, close to the respective values of 1.958 Å and 1.960 Å predicted by these authors. However, it is significantly shorter than the Cr-O distance derived from first-principles calculation (1.976 Å, Milman et al. 2001). The Cr-Mg / Cr-Si distance (3.240 (2) Å) determined from the present Rietveld analysis is close to both the distances predicted by Novak and Gibbs (1971) and Ottonello et al. (1996), i.e. 3.253 Å and 3.243 Å, respectively, and the theoretical one derived from first-principles calculation (3.255 Å).

**EXAFS experiments**

Chromium K-edge $k^3$-weighted EXAFS signal for the synthetic sample is shown in Fig. 2a, and the corresponding Fourier transform is displayed in Fig. 2b. Table 5 lists the results of the fit performed on the $k^3\chi(k)$ EXAFS data. The first neighbor contribution was fitted with 4.8 oxygen atoms at 1.96 ± 0.02 Å, which corresponds to the Cr-O distance in knorringite, in good agreement with the refined one. No attempt was made to fit it with an additional contribution of the Cr-O scattering paths in eskolaïte, because of the low proportion of this minority phase in the synthetic samples. Moreover, the Cr-O mean distance in eskolaïte (1.99 Å) is too close to that in knorringite to be resolved by EXAFS analysis.
The second peak observed in the Fourier transform (Fig. 2b) between 2.1-3.5 Å is due to a mixed contribution of Cr-Si/Cr-Mg single scattering paths in knorringite and Cr-Cr paths in eskolaïte. The fitted Cr-Si/Cr-Mg distance is 3.25 Å ± 0.04 Å, which is in good agreement with the value obtained by Rietveld refinement. The Cr-Cr distances in eskolaïte were fit to 2.94 Å ± 0.04 Å and 3.69 Å ± 0.04 Å, which are larger than those derived from the structure refinement by Newnham and deHaan (1962), i.e., 2.89 Å and 3.65 Å, respectively. This discrepancy, and the fact that the number of neighbors obtained by our EXAFS analysis is underestimated in both structures, can be likely explained by the significant overlap of the different contributions. The third peak observed on the Fourier transform between 4.1-5.3 Å is due to Cr-Cr and Cr-Si/Cr-Mg pair correlations in knorringite, at 5.01 ± 0.04 Å and 5.23 ± 0.04 Å, respectively. The fit distances are again similar to the refined ones (5.020(2) Å and 5.225(2) Å). Thus, we can conclude that the results of our EXAFS experiments are consistent with the interatomic distances obtained from Rietveld refinement of knorringite.

**Comparison between the crystal structure of high-pressure Cr-garnet end-members**

The most common Cr-garnet end-member is uvarovite $Ca_3Cr_2Si_3O_{12}$, which can be synthesised at ambient pressure (Andrut and Wildner 2002). Apart from $Mg_3Cr_2Si_3O_{12}$, which can be obtained above approximately P=11 GPa (Irifune et al. 1982), two other Cr-garnet end-members, $Mn_3Cr_2Si_3O_{12}$ and $Fe_3Cr_2Si_3O_{12}$, were obtained from synthesis runs performed at high pressure (respectively, above P=2 GPa and 6 GPa, Fursenko 1981). The necessity to increase the synthesis pressure seems to be related to the decrease of the ionic radius of the dodecahedral divalent cation (r($Ca^{2+}$) =1.12 Å, r($Mn^{2+}$) =0.96 Å, r($Fe^{2+}$) =0.92 Å and r($Mg^{2+}$) =0.89 Å) (Fursenko 1981). Although the crystal structures of $Mn_3Cr_2Si_3O_{12}$ and $Fe_3Cr_2Si_3O_{12}$ have not been yet refined, they have been predicted (Novak and Gibbs 1971; Ottonello et al. 1996). The respective lattice parameters are in good agreement with the values measured from X-ray powder diffraction data (Fursenko 1981). As in the case of knorringite, the predicted structures are similar to the one we obtained in the present work by Rietveld refinement, we can infer that the predictions made on the structures of $Mn_3Cr_2Si_3O_{12}$ and $Fe_3Cr_2Si_3O_{12}$ also yield valuable information. Therefore, we used the predicted structures given by Ottonello et al. (1996) for these two garnets and the refined structures of $Ca_3Cr_2Si_3O_{12}$ (Andrut and Wildner 2002) and $Mg_3Cr_2Si_3O_{12}$ to better understand the effect of pressure on structural changes in these garnet series. The evolution of the Cr-O distance and the Cr-O-Si angle, together with the minimum pressure of synthesis, is plotted in Figures 3a and 3b, respectively. The Cr-O distance decreases when the synthesis pressure increases, showing a significant compression of the $CrO_6$ octahedron. Additionally, the Cr-O-Si angle decreases by 3.7 % from uvarovite to knorringite, while in the mean time, the Si-O distance decreases by 1.5 % (<Si-O>= 1.6447 Å in $Ca_3Cr_2Si_3O_{12}$ vs <Si-O>= 1.620 Å in $Mg_3Cr_2Si_3O_{12}$). This confirms that the compression of the garnet structure is achieved by the rotation of the $SiO_4$ tetrahedra, rather than by their deformation (Ungaretti et al. 1995).

## ACKNOWLEDGMENTS

The authors are very grateful to G. Gudfinnsson for help in the preparation of the high-pressure synthesis runs. These experiments were performed at the Bayerisches Geoinstitut under the EU "Research Infrastructures: Transnational Access" Programme (Contract No. 505320 (RITA) - High Pressure).

**FIGURES AND TABLES**

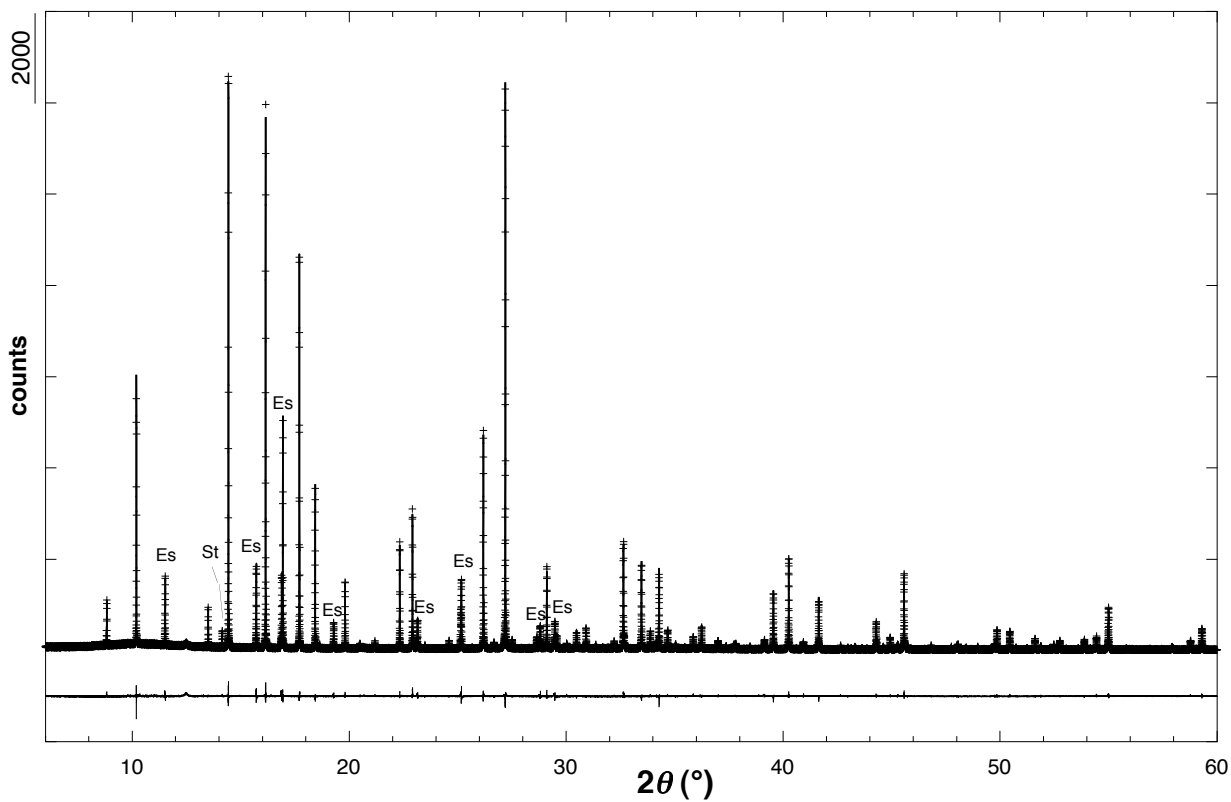

**Figure 1**. Rietveld refinement of the X-ray powder pattern of the knorringite sample. Experimental data (crosses), calculated (solid line), difference (thin solid lines). The main Bragg reflections of the eskolaite (Es) and stishovite (St) mineral impurities are indicated.

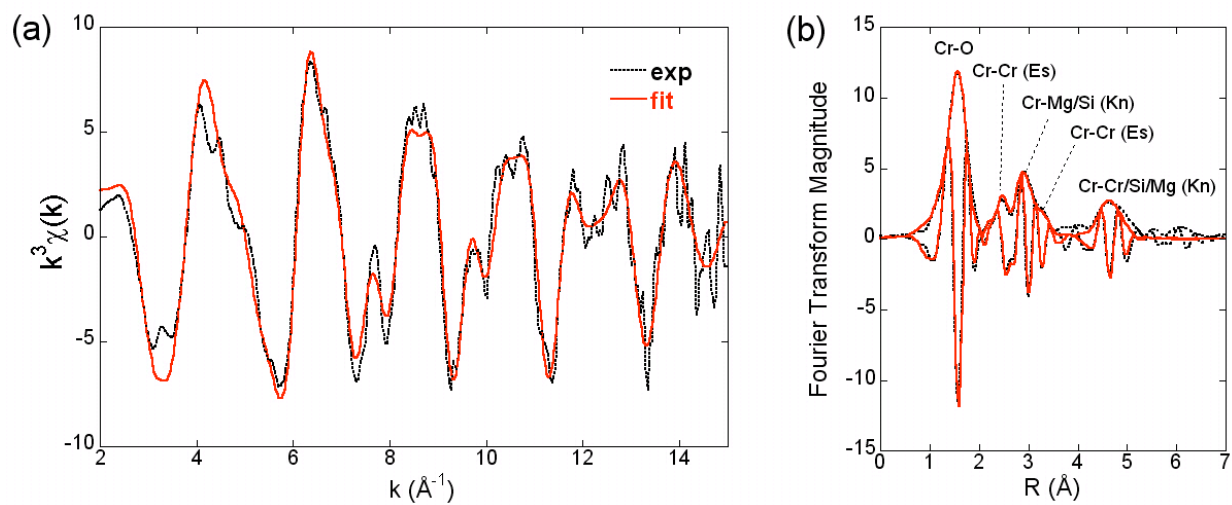

**Figure 2**. Cr K-edge unfiltered EXAFS data recorded at 293 K for synthetic knorringite sample: (a) $k^3$-weighted EXAFS, and (b) its corresponding Fourier transforms (FT), including the magnitude and imaginary part of the FT. Experimental and calculated curves are displayed as dashed and solid lines, respectively. All fit parameters are provided in Table 5.

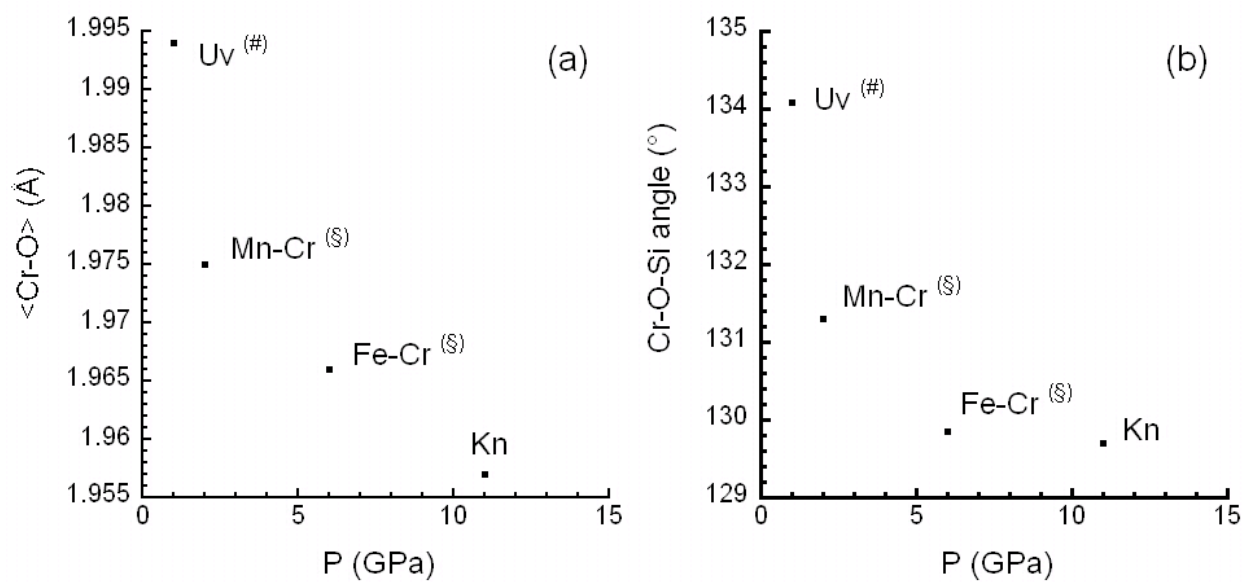

**Figure 3**. (a) Dependence of <Cr-O> distance (Å) vs minimum pressure of synthesis (GPa) for Cr-garnet end members. (b) Dependence of Cr-O-Si angle (°) vs pressure of synthesis (GPa). Uv: $Ca_3Cr_2Si_3O_{12}$, Mn-Cr: $Mn_3Cr_2Si_3O_{12}$, Fe-Cr: $Mn_3Cr_2Si_3O_{12}$ and Kn: $Mg_3Cr_2Si_3O_{12}$. #: Andrut and Wildner (2002), §: calculated from Ottonello et al (1996).

**Table 1.** Conditions and results of high-pressure multianvil runs. The following abbreviations are used : Kn : knorringite, Es : eskolaïte, St : stishovite. The run numbers are those of the diary of the Bayerisches Geoinstitut.

| Run n° | label | P (GPa) | T (°C) | Duration (h) | Appearance | Phases |
|---|---|---|---|---|---|---|
| H2732 | *Kn-1* | 11 | 1500 | 5 | Dark green | Kn + Es (+St) |
| H2733 | *Kn-2* | 14 | 1600 | 6 | Dark green | Kn + Es (+St) |

**Table 2.** Structure-refinement parameters for knorringite

| | |
|---|---|
| Formula | $Mg_3 Cr_2 (SiO_4)_3$ |
| Formula weight (g/mol) | 453.21 |
| No. of structural formula in cell unit | 8 |
| Space group | *I a -3 d* |
| $a$ (Å) | 11.5935(1) |
| Unit-cell volume, $V$ (Å$^3$) | 1558.37(4) |
| Temperature | Ambient |
| Wavelength (Å) | 0.7285 |
| Step increment ($2\theta$) | 0.0015 |
| Geometry | Debye-Scherrer |
| Background | Spline |
| Profile function | pseudo-Voigt |
| Pattern range ($2\theta$) | 6 - 60 |
| Number of reflections | 174 |
| Refined parameters | 21 |
| Rwp (%) overall | 12.4 |
| Rp (%) overall | 8.6 |
| Goodness of fit overall | 1.11 |
| $R_{Bragg}$ (%) | 3.7 |

**Table 3.** Atomic parameters for knorringite: fractional coordinates, isotropic Debye-Waller factor ($B_{iso}$), occupancy factor (Occ) and site multiplicity (Mult). Standard deviations on the last digit are under brackets.

| Atom | x/a | y/b | z/c | $B_{iso}$ (Å$^2$) | Occ | Mult |
|---|---|---|---|---|---|---|
| $Mg^{2+}$ | 1/8 | 0 | 1/4 | 1.00(8) | 1.03(1) | 24 |
| $Cr^{3+}$ | 0 | 0 | 0 | 0.31(3) | 0.93(1) | 16 |
| $Si^{4+}$ | 3/4 | 0 | 1/4 | 0.64(6) | 1.03(1) | 24 |
| $O^{2-}$ | 0.0345(2) | 0.0517(2) | 0.6569(2) | 0.76(6) | 1 | 96 |

**Table 4.** Selected distances (Å) and angles (°) in $CrO_6$/$MgO_8$/$SiO_4$ polyhedra for knorringite. Standard deviations on the last digit are under brackets.

| | Distance (Å) | | Angle (°) |
|---|---|---|---|
| Cr-O | 1.957(7) | x 6 | |
| O-O (unshared) | 2.811(7) | x 6 | 91.8(1) |
| O-O (Mg-shared) | 2.722(7) | x 6 | 88.2(1) |
| Si-O | 1.620(8) | x 4 | |
| O-O (Mg-shared) | 2.469(7) | x 2 | 99.3(1) |
| O-O (unshared) | 2.730(7) | x 4 | 114.8(1) |
| Mg-$O_a$ | 2.223(6) | x 4 | |
| Mg-$O_b$ | 2.363(6) | x 4 | |
| $O_a$-$O_a$ (Si shared) | 2.469(7) | x 2 | 67.4(1) |
| $O_a$-$O_b$ (Cr shared) | 2.722(7) | x 4 | 72.8(1) |
| $O_b$-$O_b$ (unshared) | 2.786(8) | x 2 | 72.2(1) |
| $O_a$-$O_b$ (Mg shared) | 2.695(7) | x 4 | 71.9(1) |

**Table 5.** Results of multi-shell fit of EXAFS data for the knorringite synthetic sample. ($\chi^2_{FT}$=0.38). $R$ (Å): interatomic distance, $N$: number of neighbors, $\sigma$ ( Å): Debye-Waller factor, $\Delta E_0$ (eV): difference between the user-defined threshold energy and the experimentally determined threshold energy. During the fitting procedure, all parameter values indicated by (--) were linked to the parameter value placed above in the table. Error on R value is estimated as ± 0.02 Å for the Cr-O distance and ± 0.04 Å for the other distances. Errors on $N$, $\sigma$ and $\Delta E_0$ values are estimated respectively as ± 0.5, ± 0.01 and ± 0.3, respectively.

| phase | | $R$ (Å) | $N$ | $\sigma$ (Å) | $\Delta E_0$ (eV) |
|---|---|---|---|---|---|
| Knorringite (Kn) | Cr-O | 1.96 | 4.8 | 0.062 | 4.71 |
| Knorringite (Kn) | Cr-Si | 3.25 | 2.7 | 0.032 | -- |
| Knorringite (Kn) | Cr-Cr | 5.01 | 1.7 | -- | -- |
| Knorringite (Kn) | Cr-Si | 5.23 | 2.8 | -- | -- |
| Eskolaïte (Es) | Cr-Cr | 2.94 | 0.7 | 0.065 | -- |
| Eskolaïte (Es) | Cr-Cr | 3.69 | 0.8 | -- | - |